# PERSONALIZED DATA SET FOR ANALYSIS

Vishal Gupta and Ashutosh Saxena

SETLabs, Infosys Technologies Limited, Lingampally, Hyderabad, A.P, India
vishal_gupta10@infosys.com, ashutosh_saxena01@infosys.com

## ABSTRACT

*Data Management portfolio within an organization has seen an upsurge in initiatives for compliance, security, repurposing and storage within and outside the organization. When such initiatives are being put to practice care must be taken while granting access to data repositories for analysis and mining activities. Also, initiatives such as Master Data Management, cloud computing and self service business intelligence have raised concerns in the arena of regulatory compliance and data privacy, especially when a large data set of an organization are being outsourced for testing, consolidation and data management. Here, an approach is presented where a new service layer is introduced, by data governance group, in the architecture for data management and can be used for preserving privacy of sensitive information.*

## 1. INTRODUCTION

Information requirements has now seen a renewed focus, the users are no longer willing to be provided with some static information from the consolidated warehouse alone, neither are they sporting an analyst cap with few probabilistic models to understand the customer behaviour extrapolated form the past data. The data required for today's operational and executive decision making needs to be unearthed from the large repositories of maintained or procured by the organization, such as social media; analyst's data reports and web-blogs, to name a few. The trends such as use of cloud services have been predicted [1] [2] for large data analysis and business intelligence, more so for cost of ownership and to reduce specific dependency and adopt next generation analytics solution such as in-database and in-memory. With social media analysis also taking a front row to enable a more comprehensive customer behaviour understanding and cross-sell and up-sell marketing related decision making, the data volumes in such cases are beyond the capacity of a single organization to handle or so put as not a good operational investment.

Data Management and Governance in such a complex setup becomes interestingly challenging, especially when we are seeking privacy of business sensitive information alongside a meaningful insight with consolidation of internal and external data. Today, executives no longer have to be locked in or depend on under-oath IT department to mine information out of huge chunk of data. The simplicity of solutions and boon of advanced hardware devices has given birth to the new age power users, who are capable of modeling solutions and generate information on the fly. Such information generation, in most cases, requires access to a considerable amount of business and customer sensitive information.

While cost of cyber attacks as stated by Symantec [3] mounts to around $ 2.8 million annually for large organizations, in addition organizations end up facing loss of production, revenue and worst of all customer confidence. It also states that new initiatives like cloud computing and virtualization makes security more difficult to maintain. The report also proposes that organizations' IT department should protect information proactively by protecting both

DOI: 10.5121/ijdms.2010.2404    37

information and interactions. The knowledge of where the information resides and who accesses these are very important for the Governance team to setup policies and restrict malpractices.

The internal threat due to ignorance and applications like P2P are evident in recent literature published [4], stating that sensitive information about employees and customers are available on peer-to-peer file-sharing networks and are susceptible for identity theft or fraud. Another published survey [5] cited that more than 50 percent of the organizations who provide sensitive data for project related purposes do not preserve the privacy of data before sharing. The data maybe shared outside the organization boundary for project development, testing and other knowledge outsourcing services. The requirement of data access at various groups in an organization is depicted in Figure 1.

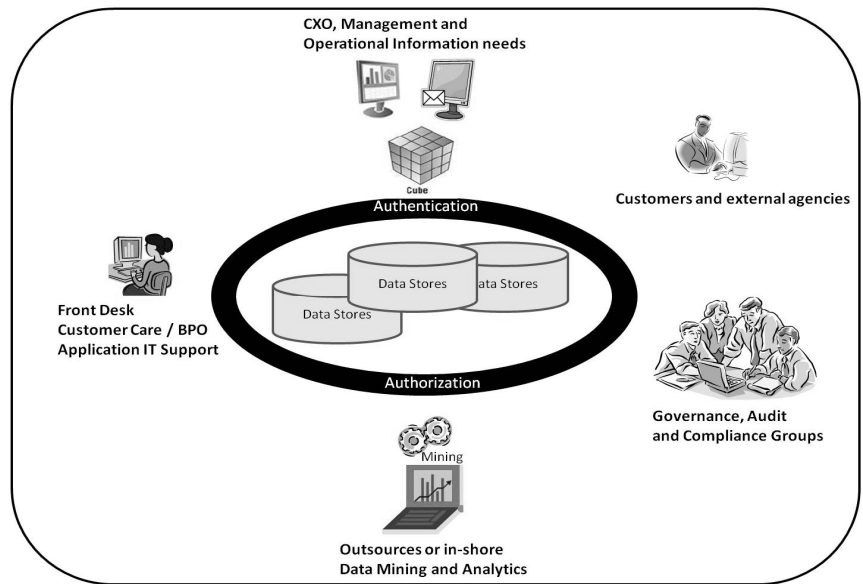

Figure 1. Data Needs for various groups of an Organization

While the data is provided to users at various levels of hierarchy, the exposure of sensitive data needs to be checked. To preserve the privacy of customer and business information, this paper presents a solution where a privacy layer is embedded as part of the organization's data architecture.

The rest of the paper is organized as follows: the section two below briefs the approaches followed to preserve privacy of data; third section provides our approach where in the constituents of the proposed privacy layer are detailed and is followed by conclusions.

## 2. PRESERVING DATA PRIVACY

Apart from being a reference for all the operational activities that happen on the day to day basis, the data stored is repurposed for additional strategic benefits such as Business Intelligence, Analytics and Mining, for details these topics one may refer [6] [7] [8] [9].The data may be shared with software service providers, third parties with whom the organization may collaborate and others such as public cloud providers for cost and technology benefits [2], eventually to enhance customer experience and reduce churn. It is required that the various levels of application development; data storage and transfer are protected such that the legal requirements of sensitive information are not compromised for the cause of strategic needs and data is protected against theft or misuse.

Some known techniques for preserving data privacy are Substitution, Transposition, Encryption, Data Masking etc. These techniques may not be ideal for application like BI, Mining and Analytics as the base level of truth is required for strategic insight. Such techniques may be useful for preparing test data and other non strategic purposes [10] [11].

An approach suggested by Peter *et al* [12] talks about alert generation on sensitive data and lets the users take the decision to manage the risks. No solution individually can support all the needs for both privacy and data mining. In large organization data may be distributed in many databases and thus the privacy in such an environment may demand a different approach, such as relational decomposition, as prescribed by Marcin and Jakub [13]. To overcome the discovery of intrusion technique reconstructing back the private information from the randomized data tuples, for a distributed environment, Zang *et al* [14] have proposed an algebraic-technique-based scheme. Ling et al [15] have proposed solution based on bloom filter to protect organizations data repurposed for business intelligence and customer private data. A weighted k-NN classification approach has been suggested by Meena *et al* [16], for mining data in the cloud environment. Synthetic data generation, which is not the same as original data, has been proposed by Vishal *et al* [17], especially for organizations who have chosen to outsource data mining tasks, as in KPO. Vassilios *et al* [18] provide a classification and description of techniques and methodologies, which have been developed under the subject of privacy preserving data mining.

For Business – to – Customer (B2C) kind of environment data perturbation techniques have been proposed, wherein the user data is initially distorted and is regenerated in a probabilistic manner to be provided to the eventual miner [19].

Nowadays analytical applications are being developed and pervasiveness is kept as a key focus area, in such scenarios the need for data access management for users in various levels of organization hierarchy becomes challenging. For preserving privacy, data replication, maintenance of additional storage for access based on role and privileges, or in a public domain is still not a profitable option for many organizations. Hence, we need a solution with minimal use of additional hardware; data duplication, user management and maintenance requirements. The solution should also not compromise the data analysis needs of users. The same solution can be used to outsource services on data without compromising on compliance requirements and data privacy.

## 3. OUR SCHEME FOR PERSONALIZED DATA SET

As observed in the techniques above, the approach proposed to preserving privacy mostly requires changing the raw data and demand additional maintenance for each new set of data that is generated or consolidated in the system. Most of the algorithms designed for preserving privacy may be unsuitable for data needs pertaining to data mining or online analytical processing. Here we propose introducing a privacy layer built in to the architecture designed for operational and large consolidated data architecture meant for strategic reporting and analysis, refer Figure 2. The approach we suggest here does not require re-engineering of the business activities and technology developed to maintain data either for operational or advance analysis. The incorporated authorization and authentication processes will also not be altered.

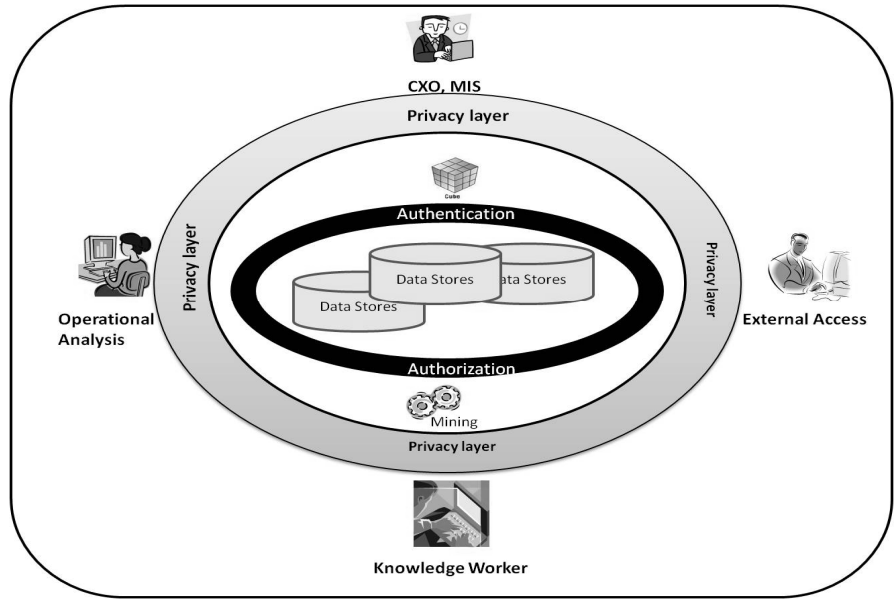

Figure 2. Architecture with Privacy Layer

The Data Source may comprise of regular online transaction processing databases, which may be used to provide near real time data for analytical processing or for any aggregated, in-memory or change data log reports; data marts and warehouses which consolidated the cleansed and transformed data from various sources, may be used for additional source for advance analytics apart from general online analytical processing. Data stores in cloud, internal or external, may also be encompassed with the privacy layer. For mining related needs the data at the root level may be directly exposed to the experts, and thus it would be required that a knowledge worker has to have access to multiple sources if the required data is not already available in a consolidated location. More of a reason for security concerns for a governance team to incorporate a privacy control on such data set requirements.

The Privacy Layer will provide a secured way for processing requests for data and also check for the privilege details of the requester. Based on the organization's governance team definitions and rules provided for the requester's role, a personalized data set is presented.

## 3.1 The Privacy Layer

The access to sensitive data would be governed by the level of the hierarchy in the organization and the current role and responsibility of the requester. Thus the access privilege and the role definition will be the key inputs for the functionality embedded in the Privacy Layer. The authorization and authentication verifications will continue to be maintained at the respective layers of the data and application architecture. This may check alone may not rule out the exposure of customer and business sensitive information requested from the consolidated data at base or any level of aggregations.

## 3.2 Components of Privacy Layer

Rule Engine and Data Range Customizer are the key components of the privacy layer, a sample data flow is shown in the Figure 3. These two components are discussed below.

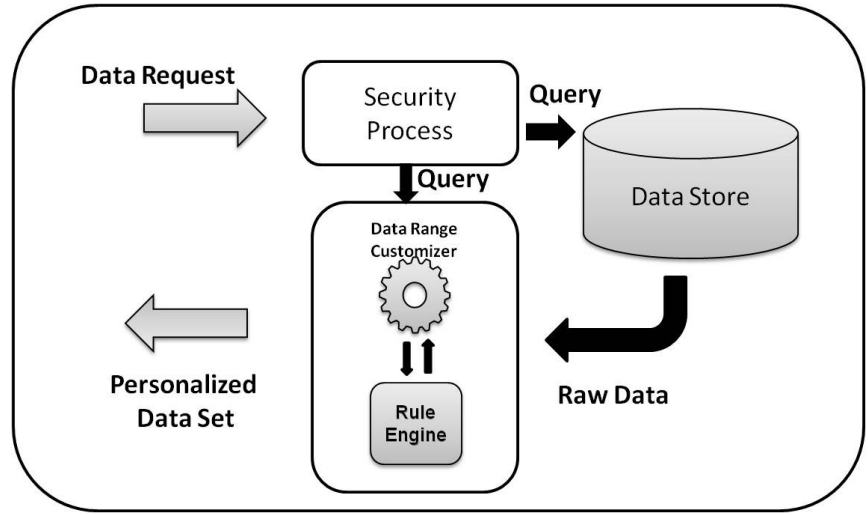

Figure 3. A sample data flow involving the privacy layer

The Rule Engine will maintain the following:

- A User Portfolio – access, role privileges, history
- Computational Algorithms – For each requester, depending upon the role and privileges, it will calculate and present personalized data set
- Rule Manager Interface – Required for the privacy administrator for maintaining the rule engine and implementing policies as laid down by governance team

The desensitizing technique utilizing the data range customizer will generate data ranges and count the occurrences of instances in each. The range for requested subject of measure for a requester will be static and will alter if his privilege and role is changed. Let us explain this with an example.

Table 1. For an actual Annual Income of customer = $75,000 per annum the following ranges will be presented to the user.

| Role | Privilege level | Data | Range $,000 |
|---|---|---|---|
| External User | Low | Age Group | 60-90 |
| Internal Operator | Medium-Low | Age Group | 65-85 |
| Managerial | Medium | Age Group | 70-80 |
| Knowledge Worker | Medium-High | Age Group | 75-80 |
| CXO | High | Age Group | 75 |

Table1 may not present the actual view of data ranges, and is a hypothetical one used for explaining the personalized data set creation technique by the privacy layer.

The length of the range provided for external user for analysis will be higher in value as compared to internal operator and manager. They, till the time they are playing the current role, will have the ranges fixed for them as provided in Table 1. Internal operator will be allocated the range 65-85 till the rule engine entries are altered for this user with a different role or changed privilege. The data sets for manager will be displaying ranges like 70-80 for fact measures and other data he has privileges to access and analyze. The knowledge worker,

41

International Journal of Database Management Systems ( IJDMS ), Vol.2, No.4, November 2010

assumed to be with higher privileges, would be having a shorter data range and CXO can fetch actual values.

Now if a conclusion from analyzed data asset in the form of a report is submitted by internal operator, based on his data range, the manager would be able to get a redistributed report for the range of data he is entitled to. When his report is viewed by the CXO, he will be able to analyze further with root level data available in the data store. The use of this approach may require changes to the way the data is presented to the user, unlike the traditional tabular format often provided by tools used to query databases. However, unlike the other techniques used for preserving data privacy, the proposed approach will not alter the raw data and will be useful in conducting a meaningful mining kind of analysis.

The establishment of a governance body and a senior management sponsorship will be a must the system to achieve its desired objective. In addition to defining the role and range sets for each requester, the security infrastructure for data level and object level will also be important to hinder unauthorized access.

The data ranges can be presented for various types of measure required for data mining and analysis. For instance:

- Item Sold: 100-500 for a given class of products
- Age Group: 25 to 35
- Sales Amount: USD 10,000 to USD 50,000
- Number of visits: 1 to 10 by a class of customers

The validity of the ranges may not clear the Chi Square test, but the point here is not to provide perfect ranges for the requested data, rather to avoid disclosure of private and sensitive business and customer information in the form of raw data and prohibit the unauthorized user to be able to regenerate the actual values.

## 4. CONCLUSIONS

In this paper an approach has been presented by us to overcome the common drawbacks available in most of the techniques used for desensitizing business and customer data. The solution proposed overcomes these drawbacks by presenting personalized data sets in the form of ranges for analysis and controlled by the role and privileges defined by the governance body of an organization.

This approach will not alter the data in the original data stores, and no additional data stores will be required for housing desensitized or synthetic data for specific set of users. Also, as the original data is not altered, there will be no requirement to maintain data integrity and hence all types of users external; operational, knowledge workers or management will be able relate their results meaningfully as it originates from the common data store, though the data sets will be controlled by a set of governing rules. The only additional requirements will be the creation of application for administrator interface, a small rule database and code development and maintenance for the data range customizer.

The code for the customizer once designed and implemented will not have to undergo changes with new data loads, and will have a minimal impact due to any strategic changes implemented by business at a process level or changes modelled within applications developed for strategic purposes. This is because, for every requester of data, who has privilege granted to access data stores will be able to analyze the data sets till his profile is maintained in the privacy layer. Additional of large volume of data in bulk will not require specific changes to the data range customizer or the rules engine. After every successful data loads to the original data stores, a mere refresh of the query will enable the requester to view the revised data set.

When the data is shared outside the legal bounds of the organization, the external user access will have a minimal impact as the data shared to him is sufficiently desensitized for him to be able to guess or derive the original values. As the root level data is restricted to a specific hierarchy in an organization, the disclosure of data, say by means peer to peer disclosure will not be difficult to track. In situations where an outsourced consultant is provided with the data sets for delivering results post analysis, he will not have any privilege to access the actual root level and his results will not vary far away from the truth.

Future work of this model is on the lines of service oriented architecture where the suggested approach of having a privacy layer will be induced as a service over the existing portfolio of information management.

This approach can be implemented within an organization for data access from data stores meant for analysis to transaction data stores. This solution can be made an integral part of enterprise data security and privacy service model, and can find place in various verticals.

## ACKNOWLEDGEMENTS

Authors would also like to acknowledge the suggestions provided by Dr. Radha Krishna Pisipati of Infosys Technologies Ltd. India. A preliminary version of the paper was presented in the First International Workshop on Trust Management in P2P Systems (IWTMP2PS 2010).

**Authors**

Vishal Gupta is a Technology Architect at SETLabs, Infosys Technologies Ltd., Hyderabad, India and received BSc Engineering (1998) and PGDBA (2000). He is certified CDMP and CBIP at Mastery Level in Data Analysis and Design (Dec 2007). As part of the consulting team of SETLabs, he is involved in assignments to provide strategic roadmap and architectural solutions for adoption of latest trends in data and information management. His interests are in the area of advanced analytics and data management solutions for building smarter organizations.

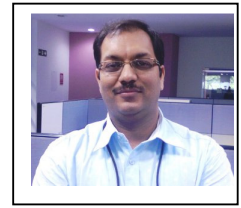

Ashutosh Saxena is a Principal Research Scientist at SETLabs, Infosys Technologies Ltd., Hyderabad, India, and received his MSc (1990), MTech (1992) and PhD in Computer Science (1999). The Indian government awarded him the post-doctorate BOYSCAST Fellowship in 2002 to research on 'Security Framework for E-Commerce' at ISRC, QUT, Brisbane, Australia. He is on the Reviewing Committees of various international journals and conferences. He has authored the book entitled *PKI – Concepts, Design and Deployment*, published by Tata McGraw-Hill, and also co-authored more than 70 research papers. His research interests are in the areas of authentication technologies, data privacy, key management and security assurance.

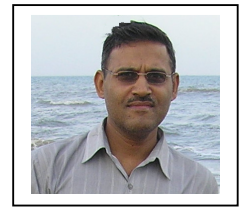